\documentclass[final,3p,times,twocolumn]{elsarticle}

\usepackage{graphicx,bm,epsf,float,amssymb}

\journal{Nuclear Instruments and Methods}

\begin{document}

\begin{frontmatter}



\title{Investigation of Large LGB Detectors for Antineutrino Detection}

\author[nps]{P.~Nelson\fnref{thanks1}}
\author[llnl]{N.~S.~Bowden\corref{cor1}}
\fntext[thanks1]{Present Address: United States Military Academy, West Point, NY~10996 USA}
\cortext[cor1]{Corresponding Author. Tel.: +1 925 422 4923.}
\ead{nbowden@llnl.gov}

\address[nps]{Department of Physics, Naval Postgraduate School, Monterey, CA~93943, USA}

\address[llnl]{Lawrence Livermore National Laboratory, Livermore, CA~94550, USA}

\begin{abstract}

A detector material or configuration that can provide an unambiguous indication of neutron capture can substantially reduce random coincidence backgrounds in antineutrino detection and capture-gated neutron spectrometry applications. Here we investigate the performance of such a material, a composite of plastic scintillator and $^6$Li$_6^{nat}$Gd$(^{10}$BO$_{3})_{3}$:Ce (LGB) crystal shards of $\approx1$~mm dimension and comprising $1\%$ of the detector by mass. While it is found that the optical propagation properties of this material as currently fabricated are only marginally acceptable for antineutrino detection, its neutron capture identification ability is encouraging.

\end{abstract}


\end{frontmatter}

\section{Introduction}
\label{sec:intro}

Neutron detection systems incorporating a neutron capture agent or neutron capture indicating detector have many applications. These include thermal neutron detection (e.g. $^3$He tubes), capture-gated neutron spectrometry~~\cite{capture}, and reactor antineutrino detection~\cite{bugey3,SONGS1}. Materials or systems that provide an unambiguous indication of a neutron capture are preferred in all these applications, as this provides a means of rejecting background. In all of the applications listed above, gamma ray rejection is of high importance, while in antineutrino detection rejection of multiple neutrons from a single cosmic ray could also be advantageous.

Here we describe an investigation of a detector material that can identify neutron captures and therefore reject a large fraction of these background events. The material, produced by MSI/Photogenics, consists of $\approx1$~mm shards of an inorganic scintillator distributed in a plastic scintillator matrix, loaded to 1\% by weight. The inorganic scintillator $^6$Li$_6^{nat}$Gd$(^{10}$BO$_{3})_{3}$:Ce (``LGB''), has a very high neutron capture cross section, high light output, and has an index of refection well matched to that of plastic scintillator~\cite{LGB1}. Neutron captures on $^6$Li or $^{10}$B are relatively easy to identify via Pulse Shape Discrimination (PSD) techniques, since the resulting heavy ions are fully contained within the crystal shards and the inorganic scintillator has a relatively long ($\approx200$~ns) decay time, compared to the organic scintillator ($\approx3$~ns). 

Previous studies of this composite material have used higher crystal loadings so as to maximize neutron capture efficiency~\cite{LGB2,LGB3}. These higher loadings led to relatively poor optical properties, and therefore limited the useable volume. Here, we consider the effect of more modest loadings, and assess the suitability of the inhomogeneous material for larger detectors like those required for for antineutrino detection and large area fast and thermal neutron detectors. 

\section{Detector Descriptions}

Two detectors were acquired from MSI/Photogenics for this investigation. The first was a cylinder of $12$~cm diameter and $12.3$~cm length containing crystal shards ranging in linear dimension between $0.5$~mm and $1.5$~mm. The second was a cylinder of $12$~cm diameter and $34.8$~cm length containing crystal shards ranging in linear dimension between $1.5$~mm and $3.0$~mm. 

\begin{figure}[tb]
\centering
\includegraphics*[width=3in]{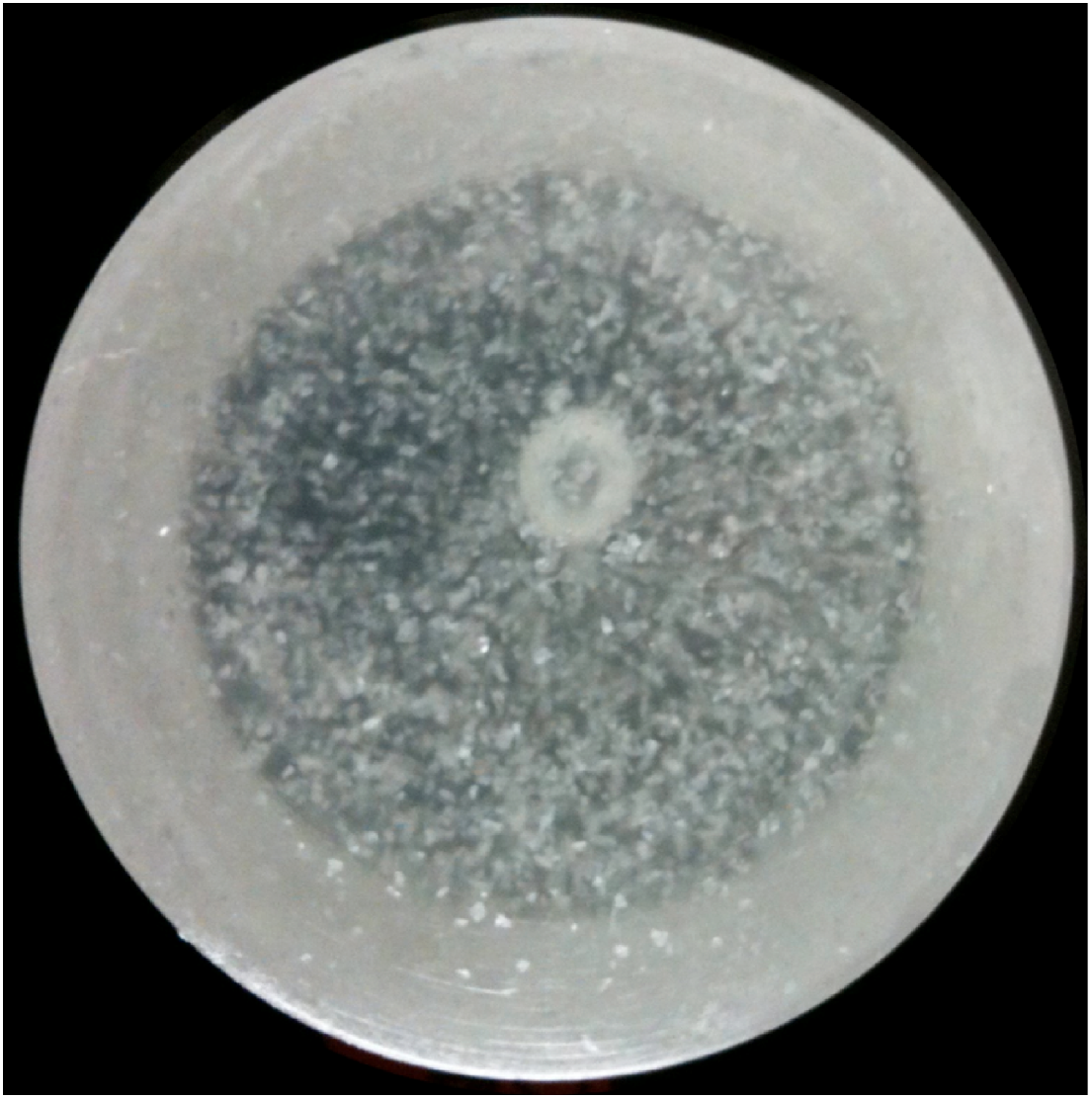}
\caption{A photograph of the smaller of the two detectors examined. The LGB crystal shards can be clealry, seen, as can a trapped air bubble.} \label{fig:photo}
\end{figure}

Each detector was loaded with $1\%$ by weight of LGB crystal shards. A photograph of the smaller detector is shown in Fig.~\ref{fig:photo}. Clearly, the optical transmission properties of the base plastic material in the visible light spectrum are affected by the inclusion of the crystal shards and a trapped air bubble is apparent. Hamamatsu H6527 $12.5$~cm Photomultipler Tube (PMT) assemblies were coupled to each end of the detector cylinders using optical grease. 

\section{Data Acquisition System}
\label{sec:daq}

\begin{figure}[tb]
\centering
\includegraphics*[width=3in]{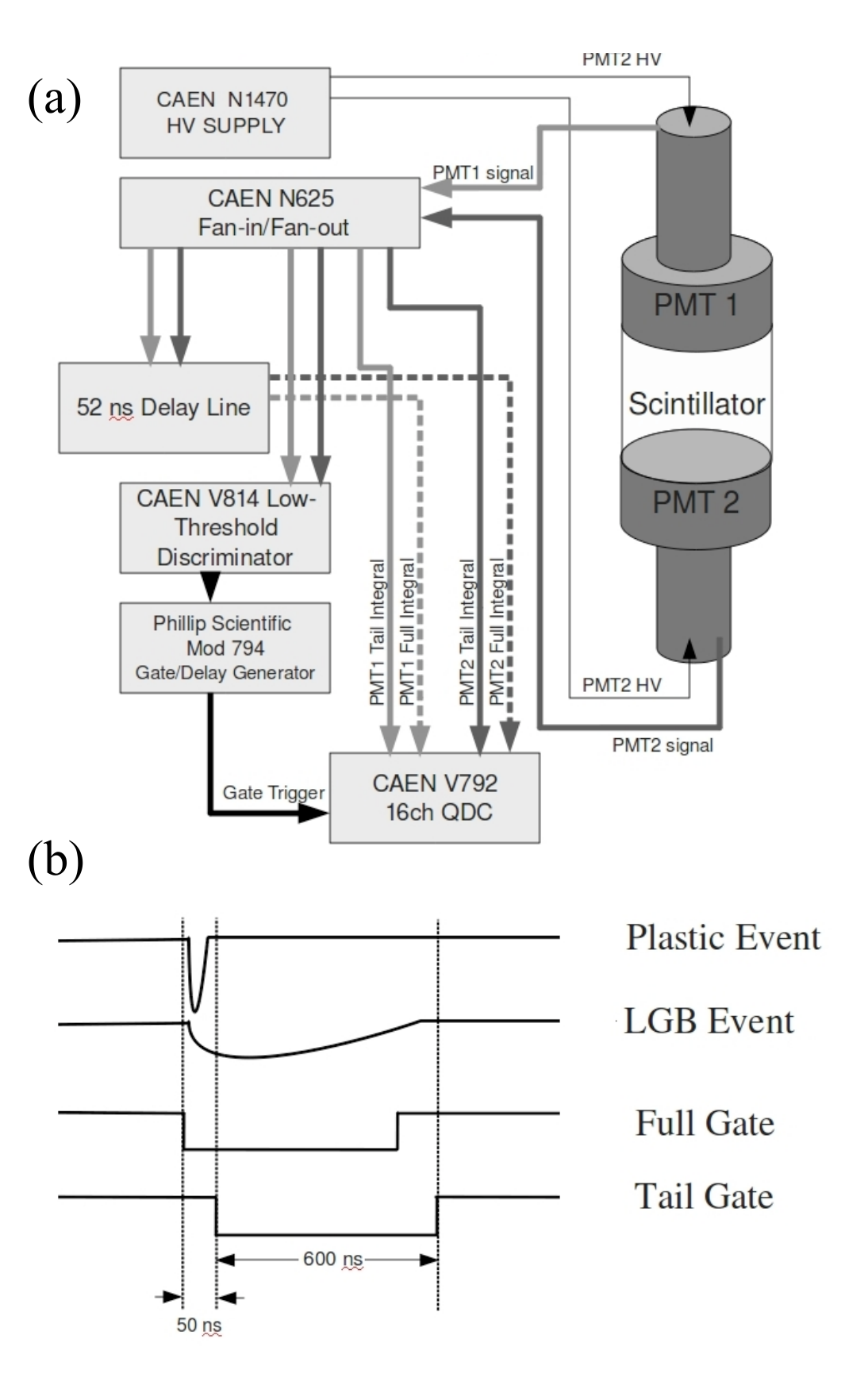}
\caption{ (a) A schematic diagram of the QDC-based DAQ system. Two copies of the PMT signal, with a relative delay, are integrated using a CAEN V792N QDC. (b) A schematic timing diagram for the QDC DAQ. Two copies of the PMTs pulses are input to the QDC with a relative delay of~$\approx50$~ns.} \label{fig:psd_daq}
\end{figure}



A schematic of the QDC-based system is shown in Fig.~\ref{fig:psd_daq}a. Two copies of each PMT signal are integrated by a CAEN V792N QDC. The delay between these two copies is carefully adjusted with respect to the QDC gate so that one copy is fully integrated, while only the tail  portion of the second copy is. As demonstrated in Fig.~\ref{fig:psd_daq}b this approach yields a different value for the ratio of integrated amplitudes for events with different decay constants. The full charge integral can be used for an event energy determination. This method is equivalent to implementing two different QDC gates during the charge integration of each PMT pulse. The ratio of the amplitudes thus determined (``tail/full'') depends upon the decay time constant of the PMT pulse. For events occurring within the fast decaying plastic scintillator this ratio will be small, whereas for events occurring within the slow decaying LGB crystals it will be large. 

%

\section{Detector Calibration}
\label{sec:calibration}

\begin{figure}[tb]
\centering
\includegraphics*[width=3in]{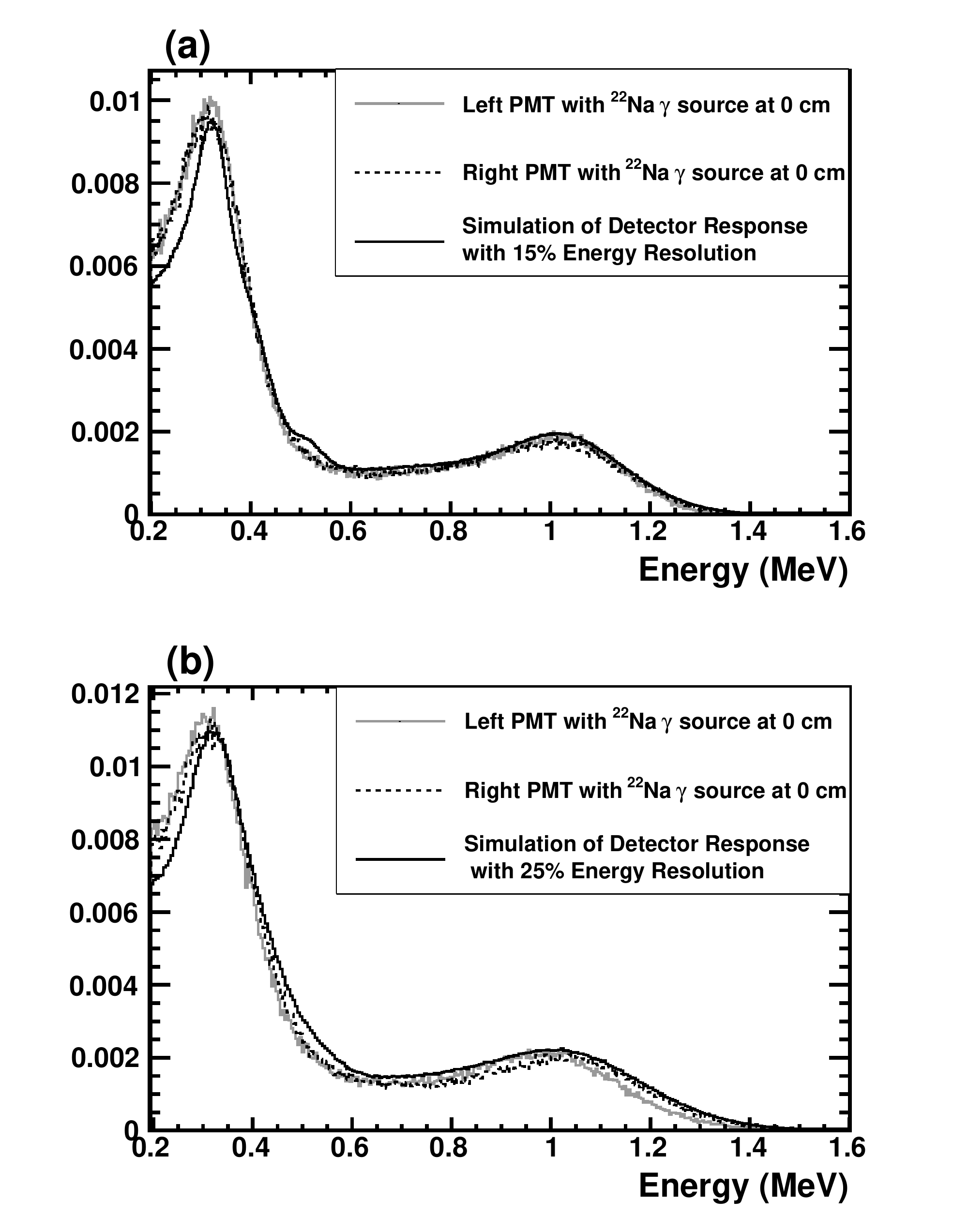}
\caption{Calibration spectra taken with a collimated $^{22}$Na source placed at the center of the  (a) $12.3$~cm and (b) $34.8$~cm detectors is compared to the simulated response. } \label{fig:cal}
\end{figure}

A relationship between recorded QDC charge and electron-equivalent energy deposition in the plastic scintillator component of the two detectors was established using the $511$~keV and $1275$~keV gamma rays emitted by a $^{22}$Na source. A ``fan'' collimator of lead bricks separated by a $0.4$~cm gap was used to preferentially illuminate the center portion of each detector. The ``Full'' response was recorded and is plotted for each PMT and detector in Fig.~\ref{fig:cal}. The energy scale was established using a GEANT4~\cite{geant} simulation of the calibration configuration - two features corresponding to the Compton edges of the two gamma rays can be clearly observed in Fig.~\ref{fig:cal} . Detector resolutions  of $15\%$ and $25\%$ were found, qualitatively, to be a good representation of the properties of the $12.3$~cm and $34.8$~cm detectors, respectively.

Using the reasonable approximation of exponential attenuation of light as it propagates through the detectors, we can combine the recorded signals of the ``left'' and ``right'' PMTs ($E_L$and $E_R$) to correct for interaction position dependencies in the analysis that follows. Assuming an effective attenuation length ($\alpha_{eff}$), that incorporates the effect of the both non-perfect reflection at the detector boundaries and optical absorption within the detector, we can write the event energy ($E$) as:

\begin{equation}
\sqrt{E_L E_R} = \sqrt{(E e^{-x/\alpha_{eff}})(E e^{x/\alpha_{eff}})} = E.
\label{eq:combine}
\end{equation}

\begin{figure}[tb]
\centering
\includegraphics*[width=3in,angle=90]{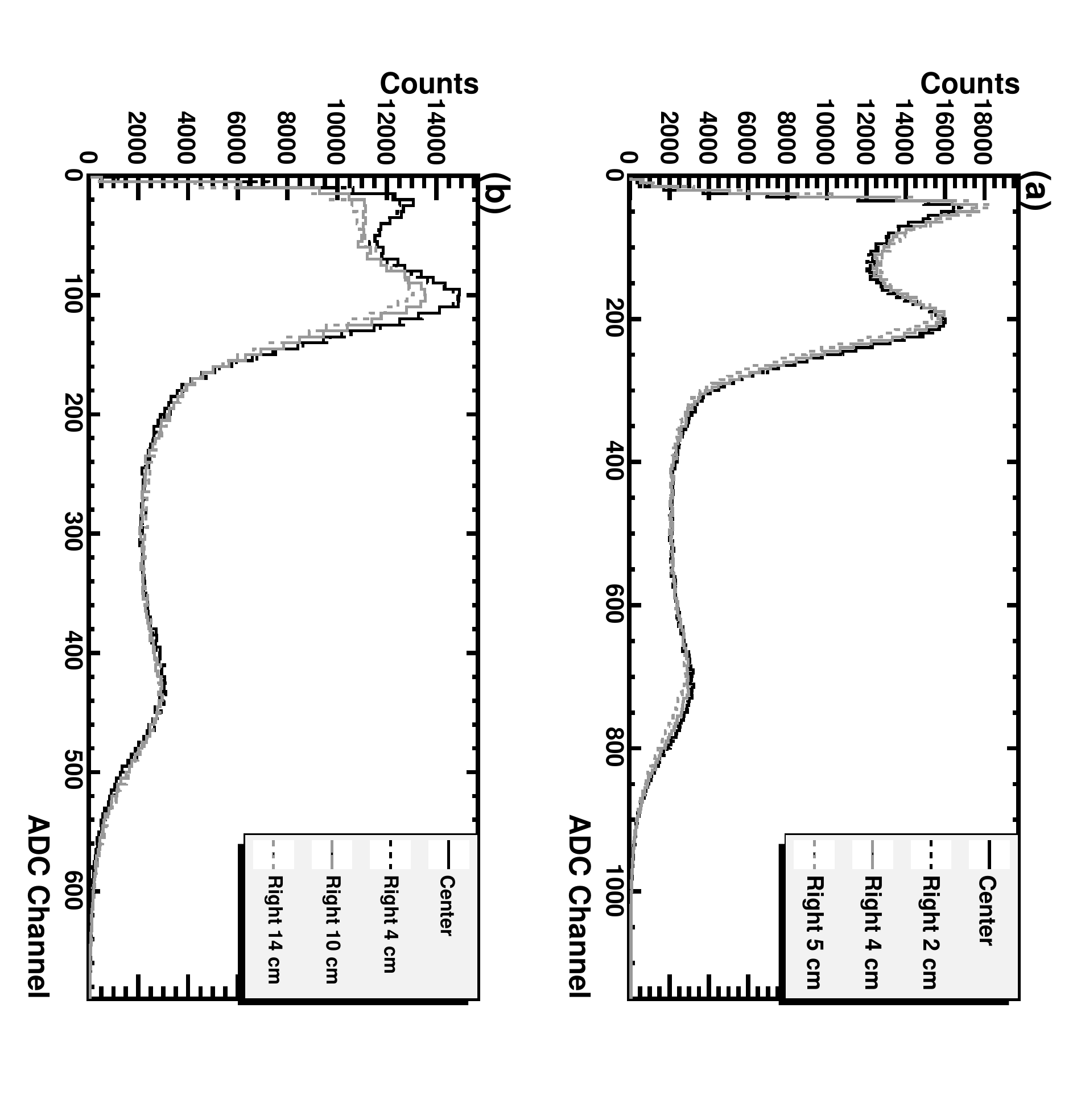}
\caption{Spectra taken with a collimated $^{22}$Na source placed at various positions relative to the center of the  (a) $12.3$~cm and (b) $34.8$~cm detectors. The good agreement between these spectra indicates that position dependencies are largely accounted for by the appropriate combination of the two PMT amplitudes.} \label{fig:position}
\end{figure}

That this is a reasonable approach here can be seen in Fig.~\ref{fig:position}, where $^{22}$Na spectra taken at several collimator positions are compared. Each is in good qualitative agreement with that taken at the center (the energy calibration position).

\section{Optical Attenuation Measurements}
\label{sec:atten}

\begin{figure}[tb]
\centering
\includegraphics*[width=3in,angle=90]{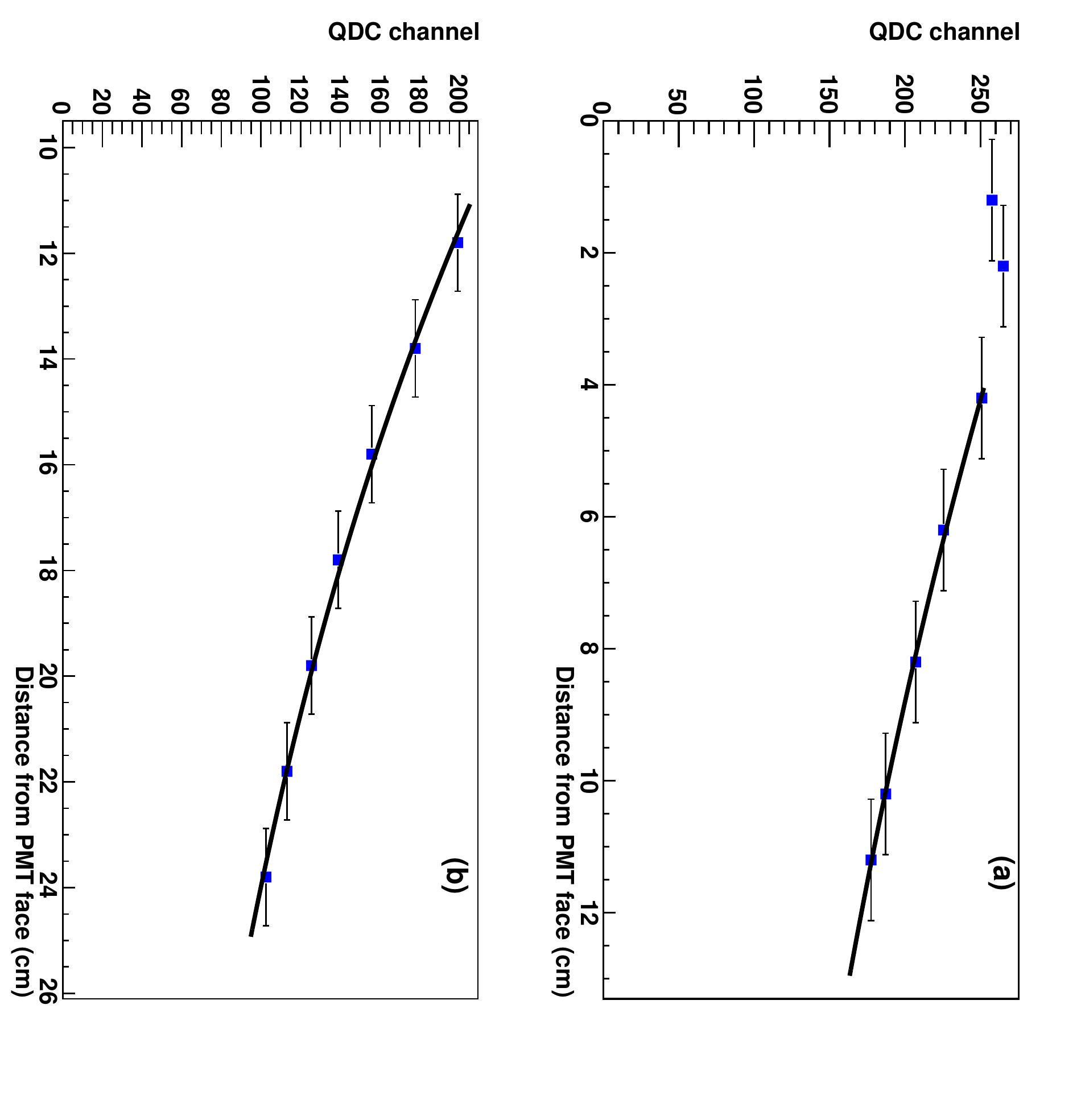}
\caption{Relative optical intensity of the Compton edge due $511$~keV $\gamma$-rays as a fucntion of $\gamma$-ray source position. Data are shown for the $12.3$~cm (a) and $34.8$~cm (b) detectors.} \label{fig:atten}
\end{figure}

The collimated gamma ray source used for energy calibration was also used to measure the effective optical attenuation length for each detector. The variation in the absolute position (measured in QDC channels) of the spectral features due to the $511$~keV and $1.275$~MeV $\gamma$-rays was recorded for each PMT from various source positions. Representative results for the ``left'' PMT of both detectors are shown in Fig~\ref{fig:atten} for the $511$~keV feature. The inferred optical attenuation lengths thus measured are given in Table.~\ref{tab:atten}. The two independent measurements made for each detector are in good agreement. 

\begin{table}
\begin{center}
\begin{tabular}{l r r} \hline
$\gamma$-ray energy&\multicolumn{2}{c}{Effective Attenuation Length (cm)}\\
&  $12.3$~cm detector & $34.8$~cm detector\\
\hline
$511$~keV  &  $17.7 \pm 2.0$~cm   &  $18.4 \pm 1.2$~cm\\
$1275$~keV  &  $16.8 \pm 2.0$~cm    &  $18.0 \pm 1.2$~cm\\
\hline\end{tabular}
\caption{\label{tab:atten} Effective attenuation lengths measured for both detectors at two $\gamma$-ray energies.}
\end{center}
\end{table}

The modest attenuation lengths measured suggest that this material, as currently manufactured, would not support detector lengths at the $1$~m scale ideal for antineutrino detection applications. In a $1$~m length detector, light from an interaction at one end of the detector would be attenuated by least 2 orders of magnitude before reaching the PMT at the other end. This would make position reconstruction and correction unreliable. However, lengths of $\approx 50$~cm appear feasible, and might be sufficient, given the attractive neutron capture identification properties of the material. 

\section{Neutron Capture Identification Results}
\label{sec:nc}

\begin{figure}[tb]
\centering
\includegraphics*[width=3in,angle=90]{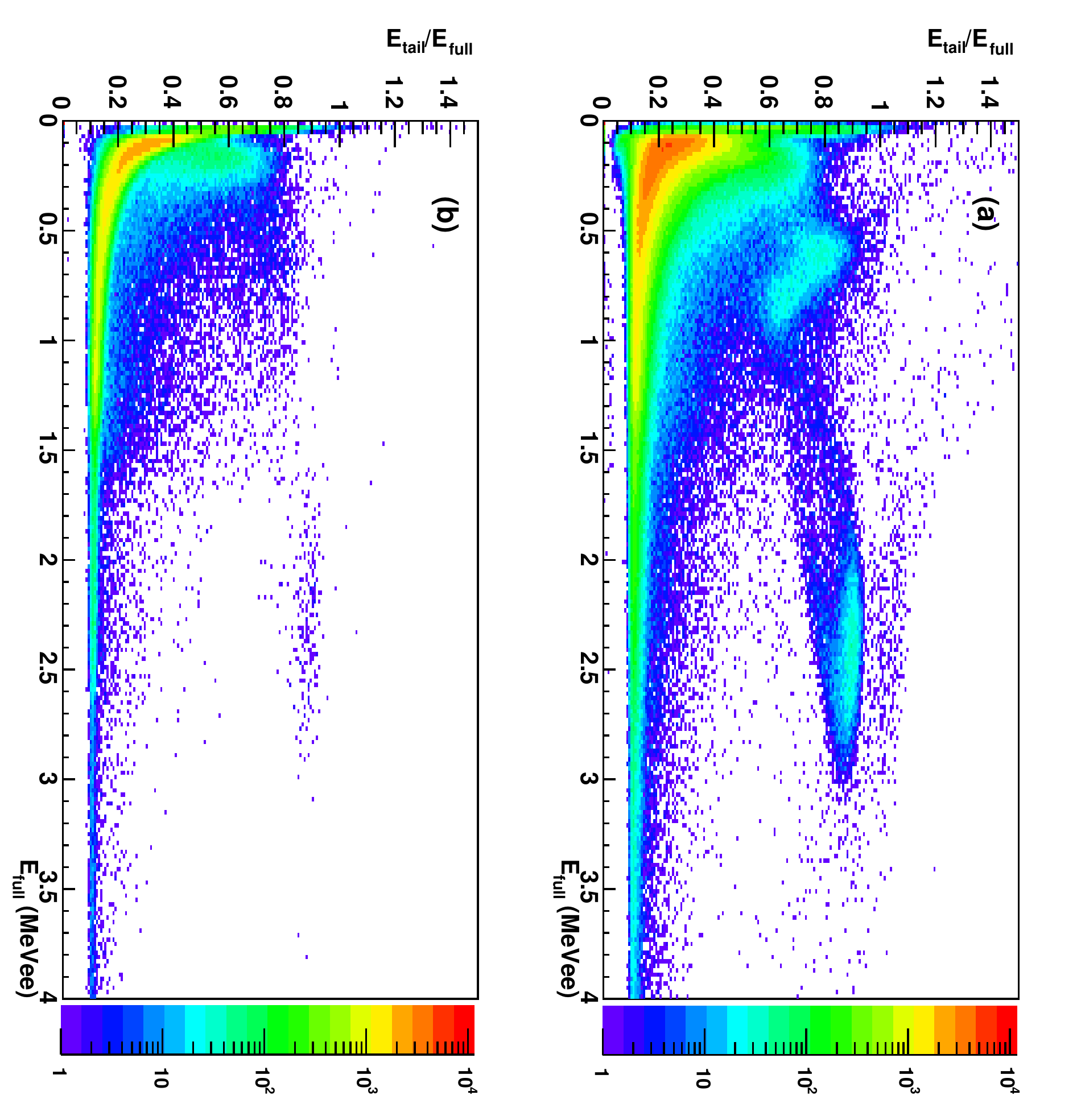}
\caption{(a) Data demonstrating the ability of the 12.3cm detector to identify neutron captures via PSD when irradiated by a $^{252}$Cf source. (b) A background spectra is shown for comparison.} \label{fig:nc_short}
\end{figure}

The neutron capture response of both detectors was investigated using a bare 2.5~$\mu$Ci~$^{252}$Cf source placed $30$~cm from the detector center. Plots of electron-equivalent energy aginst the PSD parameter (``tail'' energy/``full'' energy) from $1$~hour acquisitions are displayed in Fig.~\ref{fig:nc_short} for the $12.3$~cm detector and in  Fig.~\ref{fig:nc_short} for the $34.8$~cm detector. A background acquisition is also shown for comparison. 

\begin{figure}[tb]
\centering
\includegraphics*[width=3in,angle=90]{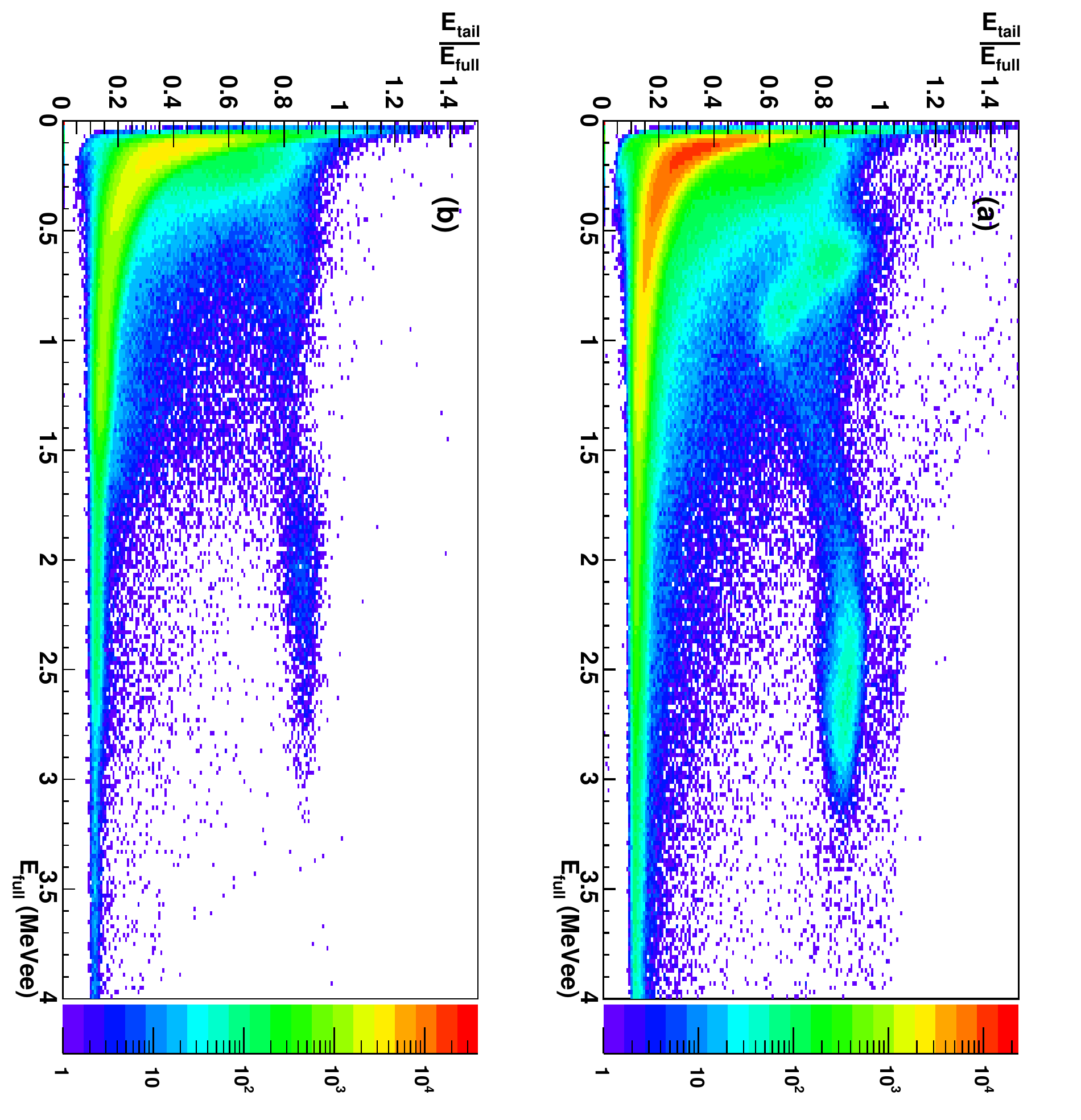}
\caption{(a) Data demonstrating the ability of the 34.8cm detector to identify neutron captures via PSD when irradiated by a $^{252}$Cf source. (b) A background spectra is shown for comparison.} \label{fig:nc_long}
\end{figure} 

Two distinct features with PSD parameters close to unity due to neutron capture can be observed in these plots: one near an electron equivalent energy of $\approx 2.2$~MeVee due to captures on $^6$Li, and the other near $0.75$~MeVee due to captures on $^{10}$B. Good separation from events in the plastic scintillator is especially evident for the $^6$Li captures. The $^{10}$B capture feature spans a greater range of values in the PSD parameter, since $93\%$ of these events are accompanied by a $478$~keV $\gamma$~ray. Compton scatters of those $\gamma$~rays add to the energy of the capture event and reduce the PSD parameter value, since they increase the ``Full'' integral value, but not that of the ``tail'' integral. 

The $^{10}$B capture feature is also more difficult to distinguish from background $\gamma$~ray interactions due to its lower average energy, which is due both to the lower Q-value of the neutron capture reaction and the higher quenching of the heavier daughter products. Low energy Compton electrons can deposit a significant fraction of their total energy in a crystal grain, raising the measured value of the PSD parameter for those events. This effect appears to contribute a significant non-neutron background to the $^{10}$B capture region.

Measurements of the neutron capture efficiency were compared to a GEANT4 simulation (Table~\ref{tab:neff}). The GEANT4 model incorporated randomly distributed LGB inclusions matching the average size quoted by the manufacturer and tallied the number of neutrons captured upon $^6$Li and $^{10}$B for an incident fission neutron spectrum. The error on the simulated efficiency was estimated by both increasing and decreasing the LGB particle size by $50\%$, since the particle size distribution is unknown. The $^{10}$B efficiency is not considered, since there are several reasons why the measured efficiency is unreliable. First, the relatively poor $\gamma$~ray discrimination of the $^{10}$B capture feature and the multitude of high energy $\gamma$~rays emitted by $^{252}$Cf made reliable neutron counting difficult with this nucleus. Second, it was noticed via inspection of oscilloscope traces that many apparent $^{10}$B neutron captures did not trigger the DAQ system. This is because the triggering is based upon a voltage level discriminator - the maximum voltage of the slow $^{10}$B PMT pulse is considerably lower than that of a $\gamma$~ray with an equivalent integrated PMT amplitude. 

\begin{table}
\begin{center}
\begin{tabular}{l r r} \hline
&\multicolumn{2}{c}{$^6$Li Capture Efficiency (\%)}\\
&  $12.3$~cm detector & $34.8$~cm detector\\
\hline
Measured&$1.71\pm0.24$&$1.38\pm0.14$\\
Simulated&$1.47\pm0.25$&$1.76\pm0.25$\\
\hline\end{tabular}
\caption{\label{tab:neff} Measured and simulated neutron capture efficiencies upon $^6$Li.}
\end{center}
\end{table}

The reasonable agreement between experiment and simulation allows us to use that same simulation to estimate the neutron capture efficiency of this material for lower neutron energies ($\approx10$~keV) relevant to the antineutrino detection application, and to investigate the effect of varying the LGB crystal isotopic composition. The GEANT4 simulation was repeated for a $1$~m$^3$ detector with $10$~keV neutrons distributed uniformly within the detector volume. The fraction of neutrons that capture on $^6$Li, $^{10}$B, Gd ($^{155}$Gd or $^{157}$Gd), H, or that escape the detector are given in Table~\ref{tab:nubar_eff} for several different LGB isotopic compositions. 

\begin{table}
\begin{center}
\begin{tabular}{l r r r r r} \hline
Isotopics&\multicolumn{5}{c}{Fraction of neutrons (\%)}\\
&$^6$Li&$^{10}$B&Gd&H&Escape \\
\hline
$^6$Li$^{nat}$Gd$^{10}$B&10.8&21.4&43.0&12.0&12.8\\
$^6$Li$^{dep}$Gd$^{10}$B&23.4&49.4&-&14.6&12.6\\
$^6$Li$^{dep}$Gd$^{11}$B&63.3&-&-&21.5&15.2\\
$^6$Li$^{nat}$Gd$^{11}$B&15.0&-&59.9&12.5&12.6\\
\hline\end{tabular}
\caption{\label{tab:nubar_eff} The predicted fraction of neutrons that capture on particular nuclei or that escape a $1$~m$^3$ detector. $^{dep}$Gd refers to gadolinium that has been completely depleted of $^{155}$Gd and $^{157}$Gd.}
\end{center}
\end{table}

Assuming that a means could be found to reliably register all $^{10}$B captures, the isotopic composition used in this investigation would yield a total neutron capture efficiency for inverse beta-decay detection of $\approx 30\%$, similar to that of previously demonstrated antineutrino detectors for reactor monitoring~\cite{SONGS1}, but far from ideal. If it were economically feasible to produce gadolinium depleted in $^{155}$Gd and $^{157}$Gd, very attractive capture efficiencies could be achieved. In particular, use of $^{dep}$Gd and $^{11}$B would be ideal, providing high efficiency for the easily identified $^6$Li capture.

\section{Conclusion}
\label{sec:conclusion}

The composite LGB/Plastic scintillator material has many attractive features for antineutrino detection. The very good PSD separation of $^6$Li neutron captures would provide a powerful means to reject random $\gamma$~ray coincidences.  This in turn might allow for a reduction in passive shielding and a reduction in total device size---an important parameter for the reactor monitoring application. That that material is based upon solid plastic could have practical advantages over the more typically used liquid scintillator.

However, this material, as currently realized, also has significant drawbacks. The optical properties do not support detector elements of the desired length. Also, despite its very high neutron capture cross section, the incorporation of gadolinium in the inorganic crystal results in a reduced identifiable neutron capture efficiency. Further work to improve the  manufacturing process that incorporates the crystals shards in the plastic would be beneficial.

Other avenues for future investigation include studying the effect of crystal shard size upon neutron capture efficiency and attempting to identify other suitable inorganic crystals that contain only $^6$Li as a neutron capture agent. Finally, we note that this material would be ideal for use in a segmented capture-gated neutron spectrometer: appropriate segmentation would allow for enhanced event-by-event resolution~\cite{seg_n}, while the unambiguous neutron capture indication would provide strong background rejection.

\section*{Acknowledgements}

LLNL-JRNL-482515.

This work was performed under the auspices of the U.S. Department of Energy by Lawrence Livermore National Laboratory in part under Contract W-7405-Eng-48 and in part under Contract DE-AC52-07NA27344.

\end{document}